\documentclass[12pt]{article}
\usepackage{amsfonts}
\usepackage{latexsym}
\usepackage{amsmath}
\usepackage{amssymb}
\usepackage{amssymb}

\newcommand{\newsection}{    
\setcounter{equation}{0}\section}
\def\appendix#1{\addtocounter{section}{1}\setcounter{equation}{0}
\renewcommand{\thesection}{\Alph{section}}
\section*{Appendix \thesection\protect\indent \parbox[t]{11.15cm}{#1}}
\addcontentsline{toc}{section}{Appendix \thesection\ \ \ #1}}

\newcommand{\be}{\begin{eqnarray}}
\newcommand{\ee}{\end{eqnarray}}
\newcommand{\bea}{\begin{eqnarray}}
\newcommand{\eea}{\end{eqnarray}}
\newcommand{\ba}{\begin{array}}
\newcommand{\ea}{\end{array}}
\newcommand{\nn}{\nonumber \\}

\def\cL{{\cal L}}

\def\bbe{{\bf{e}}}
\font\mybb=msbm10 at 11pt

\def\bb#1{\hbox{\mybb#1}}

\def\bR {\bb{R}}

\def\bC {\bb{C}}

\begin{document}

\begin{titlepage}
\vfill
\begin{flushright}
\end{flushright}
\vfill
\begin{center}
   \baselineskip=16pt
  {\Large\bf Gravitational Instantons and Euclidean Supersymmetry}
   \vskip 2cm
       J. B. Gutowski$^1$ 
      and W. A.  Sabra$^2$\\
 \vskip .6cm
      \begin{small}
      $^1$\textit{Department of Mathematics, King's College London.\\
      Strand, London WC2R 2LS\\United Kingdom \\
        E-mail: jan.gutowski@kcl.ac.uk}
        \end{small}\\*[.6cm]
  \begin{small}
      $^2$\textit{Centre for Advanced Mathematical Sciences and
        Physics Department, \\
        American University of Beirut, Lebanon \\
        E-mail: ws00@aub.edu.lb}
        \end{small}      
   \end{center}
\vfill
\begin{center}
\textbf{Abstract}
\end{center}
\begin{quote}
Supersymmetric instanton solutions in four dimensional Euclidean ungauged Einstein-Maxwell theory are analysed  and classified according 
to the fraction of supersymmetry they preserve, using spinorial geometry techniques.
\end{quote}
\end{titlepage}

\setcounter{section}{0}
\setcounter{subsection}{0}



\newsection{Introduction}

The study of instantons has connections with various branches of theoretical
physics as well as pure mathematics. For instance, instantons are an
important ingredient in the study of the nonperturbative regime of
non-abelian gauge theories and quantum mechanical systems \cite{instantons}.
The subject of non-abelian gauge fields is deeply linked to that of fibre
bundles in differential geometry which is best revealed in the relation
between instanton fermionic zero modes and the Atiyah-Singer index theorem 
\cite{atiyah}. In recent years, exact results in supersymmetric and string
theories were obtained using instanton solutions. Notable examples of
Yang-Mills instanton solutions are those obtained in \cite{polyakov}.
Gravitational instantons are defined as non-singular complete solutions to
the Euclidean Einstein equations of motion (with or without a cosmological
constant). Therefore in searching for instanton solutions, one looks for a
regular Ricci-flat or Einstein Riemannian manifolds. These are important in
semi-classical analysis of the yet unknown theory of quantum gravity.
Examples of gravitational instantons were found in \cite{eguchi}. Those
solutions, known by now as the Eguchi-Hanson instantons are the first
examples of the family of the Gibbons-Hawking instanton solutions 
\cite{Hawking}.

In the present work, we consider Einstein gravity with a $U(1)$ Maxwell
field, and define a gravitational instanton to be a Riemannian manifold
satisfying the Einstein-Maxwell field equations. Our approach to
finding solutions relies on solving a Killing spinor equation, i.e. we
look for solutions which admit Killing spinors. 
For the Euclidean theory, the first order equations obtained from the
Killing spinor equations, together with Maxwell equations and Bianchi
identity, imply the second order Einstein  equations.
In recent years, classifying supersymmetric solutions for supergravity
models in various dimensions has been an active area of research. The first
systematic classification for all Lorentzian metrics admitting
supercovariantly constant spinors in four-dimensional Einstein-Maxwell
theory was performed by Tod \cite{tod}. Tod's classification was performed
in the context of $N=2$, $D=4$ ungauged minimal supergravity 
\cite{freedman1977} employing the Penrose two-component spinor formalism 
\cite{penrose}. On setting all fermionic fields to zero, the vanishing of the
gravitini supersymmetry variation gives rise to the Killing spinor
equation.

In this paper, the techniques of spinorial geometry \cite{first}, based in part on \cite{lawson,
wang, harvey}, are used to find all gravitational instanton solutions
admitting fractions of supersymmetry. This method was first used in the
classification of solutions of supergravity theories in ten and eleven
dimensions \cite{first} and has also been useful in the classification
program in lower dimensions \cite{recentlower}. Spinorial geometry techniques
have also been used to construct the first systematic classification of
supersymmetric extremal black hole near-horizon geometries in 
ten-dimensional heterotic supergravity \cite{hethor}, and to classify the
supersymmetric solutions of Euclidean $N=4$ super Yang-Mills theory 
\cite{dietmar1}. We remark that an interesting class of Euclidean gravitational instanton
was found in \cite{maciej}, in which an analysis of the Killing spinor equation
was undertaken using the Penrose two-component spinor formalism. 
We recover this solution in our analysis; however a number of
possible spinor orbits were omitted in the calculation of \cite{maciej}.
The purpose of this paper is to perform a systematic analysis for all
possible spinor orbits, and also to classify all supersymmetric solutions
preserving higher proportions of supersymmetry.

The plan of the paper is as follows. In section two, we write down a Killing
spinor equation for the Euclidean Einstein-Maxwell theory, and investigate
its properties by considering the associated integrability condition. We
introduce various spinorial geometry techniques; the Killing
spinors are expressed in terms of differential forms and the action of the
Dirac matrices on the Dirac spinors is given. The  three \textquotedblleft
canonical\textquotedblright\ forms  into which a Dirac spinor may be placed, using
$Spin(4)$ gauge transformations are also described. We also prove
that all supersymmetric solutions must preserve either 4 or 8 supersymmetries.
In sections three and four  we fully classify all supersymmetric solutions
preserving $N=4$ and $N=8$ supersymmetries respectively.
We show that the $N=4$ solutions are either hyper-K\"ahler manifolds with
self-dual or anti-self-dual Maxwell field strengths, or the gravitational
instanton found in \cite{maciej}. We also
show that all $N=8$ solutions are locally isometric to $\bR^4$ or $S^2 \times H^2$.
In section five we present our conclusions.

\newsection{The Killing Spinor Equation}

Before we begin the classification of solutions of the Euclidean Einstein-Maxwell theory, we
will recall some facts about the corresponding Minkowskian Einstein-Maxwell
theory. The action of the theory can be written as

\begin{equation}
S=\int d^{4}x\sqrt{-g}\left( R-F_{\mu \nu }F^{\mu \nu }\right) .
\end{equation}
This constitutes the bosonic part of $N=2$ supergravity theory 
\cite{freedman1977}. The gravitino Killing spinor spinor
equation is

\begin{equation}
\left( \nabla _{\mu }+{\frac{i}{4}}F_{\nu _{1}\nu _{2}}\Gamma ^{\nu _{1}\nu
_{2}}\Gamma _{\mu }\right) \epsilon =0
\end{equation}
where $\nabla _{\mu }$ is the covariant derivative given by
\begin{equation}
\nabla _{\mu }=\partial _{\mu }+{\frac{1}{4}}\Omega _{\mu ,\nu _{1}\nu
_{2}}\Gamma ^{\nu _{1}\nu _{2}} \ ,
\end{equation}
$\Omega$ is the spin connection, 
and $\epsilon $ is a Dirac spinor.

For Einstein-Maxwell theory in Euclidean spacetime, we shall consider what, a priori, appears to be a more
generic Killing spinor equation:
\begin{equation}
\label{kse1}
\nabla _{\mu }\epsilon =cF_{\nu _{1}\nu _{2}}\Gamma ^{\nu _{1}\nu
_{2}}\Gamma _{\mu }\epsilon 
\end{equation}
where $c$ is a complex constant. We shall require that the integrability
conditions of this equation imply that the Einstein equations of motion hold.
The integrability conditions of ({\ref{kse1}}) imply that
\begin{equation}
\label{einst1}
R_{\mu \nu }+8c^{2}\left( 4F_{\mu \sigma }F_{\nu }{}^{\sigma }-g_{\mu \nu
}F_{\nu _{1}\nu _{2}}F^{\nu _{1}\nu _{2}}\right) =0.
\end{equation}
Hence, in general, we shall require that $c^2$ is real, so either $c$ is real or 
imaginary. However, this choice is merely an artefact of the 
representation of the Clifford algebra. To see this, one can define an equivalent representation
of the Clifford algebra $\Gamma'$ by
\be
\Gamma_\mu = i \Gamma_5 \Gamma'_\mu
\ee
where $\Gamma_5=\Gamma_{1234}$. On substituting this into the Killing spinor equation ({\ref{kse1}})
one obtains
\be
\nabla_\mu \epsilon = -ic \star F_{\nu_1 \nu_2} \Gamma'^{\nu_1 \nu_2} \Gamma'_\mu \epsilon
\ee
Hence it is clear that the solutions for which $c$ is real are the same as those for which
$c$ is imaginary, provided one replaces $F$ with its Hodge dual $\star F$.

There is however a special case, when $F$ is self-dual (or anti-self-dual). In this case, 
the contribution from the Maxwell fields in ({\ref{einst1}}) vanishes, and hence the integrability
conditions of ({\ref{kse1}}) do not constrain $c^2$ to be real any longer. We remark that in
the case of the gauged Euclidean supergravity theory, whose anti-self-dual solutions were classified in
\cite{maciejgut}, the relaxation of the reality condition on this coefficient of the Killing spinor equation, 
which occurs when $F$ is self-dual or anti-self dual, produces an enlarged class of solutions. The geometries
of these solutions depend on the phase of $c$. This is though not the case in the ungauged theory, as one can again
see that for solutions with $F= \pm \star F$, the phase of $c$ is once more simply an artefact of the representation of the
Clifford algebra. In particular, on defining $\Gamma'$ by
\be
\Gamma_\mu = (\cos \xi \mp i \sin \xi \Gamma_5) \Gamma'_\mu
\ee
for real constant $\xi$,
so that $\Gamma'$ is an equivalent representation of the Clifford algebra, one finds that
({\ref{kse1}}) can be rewritten as
\be
\nabla_\mu \epsilon = c e^{i \xi}  F_{\nu_1 \nu_2} \Gamma'^{\nu_1 \nu_2} \Gamma'_\mu \epsilon \ ,
\ee
and an appropriate choice of $\xi$ can be used to set, without loss of generality, $c=-{i \over 4}$.

So henceforth we shall set $c=-{i \over 4}$ in the Killing spinor equation ({\ref{kse1}}).

\subsection{Spinorial Geometry}

In order to apply spinorial geometry techniques to analyse solutions of the  Killing spinor equation ({\ref{kse1}}),
we define a complex \textit{spacetime} basis $\mathbf{e}^{1},
\mathbf{e}^{2},\mathbf{e}^{\bar{1}},\mathbf{e}^{\bar{2}}$, for which the spacetime
metric is 
\begin{equation}
ds^{2}=2\left( \mathbf{e}^{1}\mathbf{e}^{\bar{1}}+\mathbf{e}^{2}
\mathbf{e}^{\bar{2}}\right) .
\end{equation}
Moreover, the space of Dirac spinors is taken to be the complexified space
of forms on $\mathbb{R}^{2}$, with basis $\{1,e_{1},e_{2},e_{12}=e_{1}\wedge
e_{2}\}$; a generic Dirac spinor $\epsilon $ is a complex linear combination
of these basis elements. In this basis, the action of the Dirac matrices 
$\Gamma _{m}$ on the Dirac spinors is given by 
\begin{equation}
\Gamma _{m}=\sqrt{2}i_{e_{m}},\qquad \Gamma _{\bar{m}}=\sqrt{2}e_{m}\wedge 
\end{equation}
for $m=1,2$. We define 
\begin{equation}
\Gamma _{5}=\Gamma _{1\bar{1}2\bar{2}}
\end{equation}
which acts on spinors via 
\begin{equation}
\Gamma _{5}1=1,\qquad \Gamma _{5}e_{12}=e_{12},\qquad \Gamma
_{5}e_{m}=-e_{m}\quad m=1,2.
\end{equation}

There are three non-trivial orbits of $Spin(4)=Sp(1)\times Sp(1)$ acting on
the space of Dirac spinors \cite{bryant}. In our notation, and following the
reasoning set out in \cite{half2007}, one can use $SU(2)$ transformations to
rotate a generic spinor $\epsilon $ into the canonical form 
\begin{equation}
\epsilon =\lambda 1 +\sigma e_{1}
\end{equation}
where $\lambda ,\sigma \in \mathbb{R}$. The three orbits mentioned above
correspond to the cases $\lambda =0$, $\sigma \neq 0$; $\lambda \neq 0$, 
$\sigma =0$ and $\lambda \neq 0$, $\sigma \neq 0$. The orbits corresponding
to $\lambda =0$, $\sigma \neq 0$ and $\lambda \neq 0$, $\sigma =0$ are
equivalent under the action of $Pin(4)$.

It will also be useful to define a charge conjugation operator  $C$,
which acts on spinors via
\begin{equation}
C1=-e_{12},\quad Ce_{12}=1,\quad Ce_{i}=-\epsilon _{ij}e_{j} \ .
\end{equation}
It is straightforward to show that if $\epsilon$ is a solution to the Killing spinor equation
({\ref{kse1}}), then so is
\be
\epsilon' = \Gamma_5 C* \epsilon
\ee
and moreover $\epsilon, \epsilon'$ are linearly independent over $\bC$. 
Also observe that ({\ref{kse1}}) is linear over $\bC$, so it follows that
all supersymmetric solutions of ({\ref{kse1}}) must preserve $4$, $6$ or $8$ supersymmetries.

In fact, one can straightforwardly exclude the possibility of solutions with exactly 6 supersymmetries as follows.
Suppose there is a $N=6$ solution, which is not maximally superymmetric, 
whose space of Killing spinors is spanned (over $\bC$) by
$\{ \epsilon_1, \epsilon_2, \epsilon_3 \}$. From the previous argument, one can 
set, without loss of generality $\epsilon_2= \Gamma_5 C* \epsilon_1$. 
Then $\Gamma_5 C* \epsilon_3$ is a Killing spinor. Suppose that
one can write this spinor as a complex linear combination of
of $\epsilon_1, \Gamma_5 C* \epsilon_1, \epsilon_3$, so 
\be
\label{lc1}
\Gamma_5 C* \epsilon_3 = \lambda_1 \epsilon_1 + \lambda_2 \Gamma_5 C* \epsilon_1 + \lambda_3 \epsilon_3
\ee
for $\lambda_1, \lambda_2, \lambda_3 \in \bC$. Acting on both sides with $\Gamma_5 C*$ one finds
\bea
\label{lc2}
-\epsilon_3 &=& {\bar{\lambda}}_1 \Gamma_5 C* \epsilon_1 - {\bar{\lambda}}_2 \epsilon_1
+ {\bar{\lambda_3}} \Gamma_5 C* \epsilon_3
\nn
&=& \big(- {\bar{\lambda}}_2 + \lambda_1 {\bar{\lambda}}_3 \big) \epsilon_1
+ \big({\bar{\lambda}}_1 + \lambda_2 {\bar{\lambda}}_3 \big) \Gamma_5 C* \epsilon_1 + |\lambda_3|^2 \epsilon_3
\eea 
where ({\ref{lc1}}) has been used to rewrite the $ \Gamma_5 C* \epsilon_3$
term in ({\ref{lc2}}). However, this cannot hold, as it would require $|\lambda_3|^2=-1$.
Hence, $\{ \epsilon_1, \Gamma_5 C* \epsilon_1, \epsilon_3, \Gamma_5 C* \epsilon_3 \}$
must be a linearly independent set of Killing spinors, in contradiction to the assumption that
the solution is not maximally supersymmetric. Hence, there can be no
(exactly) $N=6$ solutions. A very similar argument was used to rule out $N=6$ solutions for
the minimal five-dimensional ungauged supergravity in 
\cite{d5class}.

\newsection{$N=4$ Solutions}

In this section, we shall analyse the half-supersymmetric solutions of ({\ref{kse1}}), with
Killing spinors $\epsilon = \lambda 1 +\sigma e_1$ and $\epsilon' = \Gamma_5 C* \epsilon = \sigma e_2
-\lambda e_{12}$. The linear system obtained by substituting
the spinor $\epsilon$ into the Killing spinor equation ({\ref{kse1}}) is given in the Appendix.

\subsection{Solutions with $\epsilon= \lambda 1$ or $\epsilon = \sigma e_1$}

To begin, we analyse the solutions for which $\epsilon= \lambda 1$ with $\lambda \neq 0$.
In this case, the linear system ({\ref{linsys}}) implies that
\be
d \lambda =0
\ee
together with the following conditions on the spin connection
\begin{eqnarray}
\Omega _{\mu, 1{\bar{1}}}{}+\Omega _{\mu, {2{\bar{2}}}} &=&0,  \qquad
\Omega _{\mu,12} =0,  \label{sdual}
\end{eqnarray}
for all $\mu$, and the Maxwell field strength must satisfy
\begin{equation}
F_{1{\bar{1}}}=F_{2{\bar{2}}},\text{ \ \ \ \ \ }F_{{\bar{1}}2}=0.
\end{equation}

The conditions on the spin connection imply that the
almost complex structures $I^1, I^2, I^3$ defined by
\be
I^1= \bbe^{12}+ \bbe^{\bar{1} \bar{2}},  \qquad 
I^2 = i(\bbe^{12}+ \bbe^{\bar{1} \bar{2}}), \qquad
I^3 = i (\bbe^{1 \bar{1}}+ \bbe^{2 \bar{2}})
\ee
and which satisfy the algebra of the imaginary unit quaternions, are covariantly
constant with respect to the Levi-Civita connection, i.e. the manifold is hyper-K\"ahler.
The conditions on the Maxwell field strength imply that $F$ is anti-self-dual. 
This exhausts the content of the Killing spinor equations for these solutions.

We remark that the first order differential equations coming from (\ref{sdual})
were actually imposed in the derivations of the gravitational instanton 
\cite{eguchi}. This is in analogy with imposing the self-duality condition on the
Yang-Mills field strength in the derivation of Yang-Mills instantons 
\cite{polyakov}. In our present consideration, the self duality of the spin
connection and the associated first order differential equations are the
result of the requirement of Euclidean supersymmetry.

Similarly, the analysis of the solutions for which $\epsilon= \sigma e_1$, $\sigma \neq 0$, gives the following
conditions
\begin{eqnarray}
d \sigma &=& 0,  \notag \\
\Omega _{\mu, 1{\bar{1}}} &=&\Omega_{\mu, 2{\bar{2}}},\text{ \ \ \ }\Omega _{\mu,{\bar{1}2
}}=0  \notag \\
F_{1{\bar{1}}}+F_{2{\bar{2}}} &=&0,\text{ \ \ \ \ \ }F_{{1}2}=0.
\end{eqnarray}
Hence the conditions on the geometry and $F$ are equivalent to those found for the solutions with
$\epsilon=\lambda 1$, but with the opposite
duality for both the spin connections and the gauge field strength; i.e.
the manifold is again hyper-K\"ahler, and $F$ is self-dual.

\subsection{Solutions with $\epsilon=\lambda 1 + \sigma e_1$}

Suppose now that $\epsilon = \lambda 1+ \sigma e_1$, with $\lambda \neq 0$ and
$\sigma \neq 0$. It will be convenient to define the 1-form
\be
V = i \lambda \sigma (\bbe^1-\bbe^{\bar{1}}) \ .
\ee
On taking appropriate linear combinations of the equations in ({\ref{linsys}}),
one finds that the vector field dual to $V$ (which we also denote by $V$) is
Killing, and moreover satisfies
\be
\cL_V \lambda = \cL_V \sigma =0,  \qquad \cL_V F=0 \ .
\ee
The geometric conditions also imply that 
\be
d\left( \lambda \sigma \left( \mathbf{e}^{1}+\mathbf{e}^{\bar{1}}\right)
\right) =d\left( \lambda \sigma \mathbf{e}^{2}\right) =0 \ .
\ee
Hence, we introduce real local co-ordinates $\tau$, $x$, $y$, $z$ by
\be
V = \sqrt{2} {\partial \over \partial \tau}
\ee
and
\be
\bbe^1+\bbe^{\bar{1}} = {\sqrt{2} \over \lambda \sigma} dx, \qquad
\bbe^2 =  {1 \over \sqrt{2} \lambda \sigma}(dy+idz)
\ee
such that the 1-form $V$ is given by
\be
V = \sqrt{2} (\lambda \sigma)^2 ( d \tau + \phi )
\ee
where
\be
\phi = \phi_x dx+ \phi_y dy + \phi_z dz
\ee
is a 1-form on $\bR^3$, and $\lambda, \sigma, \phi_x, \phi_y, \phi_z$ are all independent of $\tau$.
Hence, the metric is given by
\begin{equation}
ds^{2}=\left( \lambda \sigma \right) ^{2}(d\tau +\phi )^{2}+\frac{1}{\left(
\lambda \sigma \right) ^{2}}\left( dx^{2}+dy^{2}+dz^{2}\right) \ .
\end{equation}
The remaining geometric conditions obtained from ({\ref{linsys}}) imply that
\be
d \phi = {2 \over (\lambda \sigma)^2} \star_3 d  \log \bigg( {\lambda \over \sigma} \bigg) \ .
\ee
The gauge field strength is given by
\begin{equation}
F=\frac{1}{2}(d\tau +\phi )\wedge d\left( \sigma ^{2}-\lambda ^{2}\right) 
- \frac{1}{2\left( \lambda \sigma \right) ^{2}}\ast _{3}d\left( \lambda
^{2}+\sigma ^{2}\right) \ .
\end{equation}

It remains to impose the Bianchi identity $dF=0$ and Maxwell equation $d \star F=0.$ These, respectively, give
\bea
\bigg( {\lambda^2 \over \sigma^2} \bigg) \nabla^2 \lambda^{-2} + 
\bigg( {\sigma^2 \over \lambda^2} \bigg) \nabla^2 \sigma^{-2} &=&0
\nn
\bigg( {\lambda^2 \over \sigma^2} \bigg) \nabla^2 \lambda^{-2} -
\bigg( {\sigma^2 \over \lambda^2} \bigg) \nabla^2 \sigma^{-2} &=&0
\eea
where $\nabla^2$ is the Laplacian on $\bR^3$.
It follows that both $\frac{1}{\sigma ^{2}}$ and $\frac{1}{\lambda ^{2}}$ are harmonic functions on $\bR^3$. 
This solution is identical to that found in \cite{maciej} using the Penrose two-component spinor formalism.
We remark that there are two special cases. If $\lambda$ is constant, then $F$ is self-dual,
whereas if $\sigma$ is constant then $F$ is anti-self-dual. In both these cases, the metric is
the Gibbons-Hawking solution of \cite{Hawking}.

\newsection{$N=8$ Solutions}

In order to analyse the necessary conditions for a solution to
preserve 8 supersymmetries, note that 
the integrability condition of the Killing spinor equation ({\ref{kse1}}) can
be written as 
\begin{equation}
\left( (T_{\mu \nu }^{1})_{\tau }\Gamma ^{\tau }+(S_{\mu \nu }^{1})_{\tau
}\Gamma ^{\tau }\Gamma _{5}+{\frac{1}{2}}(T_{\mu \nu }^{2})_{\lambda
_{1}\lambda _{2}}\Gamma ^{\lambda _{1}\lambda _{2}}\right) \epsilon=0,
\label{inter1}
\end{equation}
where the tensors $T^{1},S^{1},T^{2}$ are defined via 
\begin{eqnarray}
(T_{\mu \nu }^{2})_{\lambda _{1}\lambda _{2}} &=&{\frac{1}{2}}R_{\mu \nu
\lambda _{1}\lambda _{2}}
\nn
&+&{1 \over 2} \bigg( F_{\sigma _{1}\sigma _{2}}F^{\sigma
_{1}\sigma _{2}}\delta _{\mu \lbrack \lambda _{1}}\delta _{\lambda _{2}]\nu
}-2\delta _{\nu \lbrack \lambda _{1}]}F_{\mu }{}^{\sigma }F_{\sigma |\lambda
_{2}]}+2\delta _{\mu \lbrack \lambda _{1}]}F_{\nu }{}^{\sigma }F_{\sigma
|\lambda _{2}]}\bigg) ,  \notag \\
(T_{\mu \nu }^{1})_{\sigma } &=&{i \over 2}\nabla _{\sigma }F_{\mu \nu },  \notag \\
(S_{\mu \nu }^{1})_{\sigma } &=&{i \over 2}\nabla _{\sigma }\star F_{\mu \nu } \ .
\end{eqnarray}

{}For a $N=8$ solution, ({\ref{inter1}}) must hold for all choices of $\epsilon$, and hence
one must have $T^1=S^1=0$, so $F$ is covariantly constant, and also $T^2=0$.
The $N=8$ solutions are obtained by noting that one can always apply a $SO(4)$ transformation to set
\be
F = \alpha \bbe^1 \wedge \bbe^2 + \beta \bbe^3 \wedge \bbe^4
\ee
for real functions $\alpha, \beta$.
As $\nabla F=0$, this implies that $F \wedge \star F$ and $F \wedge F$ are covariantly constant, and hence
$\alpha, \beta$ are constant.
On substituting this expression for $F$ into the condition $T^2=0$, one finds that the only possible non-vanishing
components of the Riemann tensor are given by
\be
R_{3434}=-R_{1212}= 16c^2 (\alpha^2-\beta^2) \ .
\ee
Hence, if $\alpha^2 \neq \beta^2$ then the manifold is locally isometric to $S^2 \times H^2$, where $R_{S^2}= - R_{H^2} = 32 c^2 |\alpha^2-\beta^2|$, and $F$ is a linear combination of the volume forms of $S^2$ and $H^2$, which is neither self-dual or
anti-self-dual.
However, if $\alpha^2= \beta^2$ then the manifold is flat, and $F$ is either self-dual or anti-self-dual;
\be
ds^2= \delta_{ij
} dx^i dx^j, \qquad F = k (dx^{12}\pm dx^{34})
\ee
for constant $k$.

\newsection{Conclusions}

We have defined supersymmetric gravitational instantons as solutions of the
Euclidean Einstein-Maxwell equations of motion admitting Killing spinors.
Using spinorial geometry techniques, similar to those which have been 
particularly successful in classifying solutions of Lorentzian supergravity theories
in higher dimensions, we have fully classified these solutions. By making
use of the gauge freedom of the Killing spinor equation, one is able to
reduce a generic spinor to one of three canonical forms.
We have proven that all supersymmetric solutions are either half-supersymmetric
or maximally supersymmetric.
Solutions admitting half of the supersymmetry are given in terms of 
hyper-K\"{a}hler manifolds with self dual or anti-self dual gauge field strengths,
and the solution found by Dunajski and Hartnoll.
The maximally supersymmetric solutions are locally isometric either to
$\bR^4$ or to $S^2 \times H^2$.
Our analysis in this paper can be extended to cases with a non-vanishing
cosmological constant and to theories with scalar fields. Work along these
directions is in progress.

\vskip 0.5cm
{\bf Acknowledgments:}~The work of WS was supported in part by the National Science
Foundation under grant number PHY-0903134. 
JG is supported by the EPSRC grant EP/F069774/1.

\setcounter{section}{0}
\setcounter{subsection}{0}

\appendix{The Linear System}

In this Appendix, we list the linear system of equations obtained from
the Killing spinor equation ({\ref{kse1}}), with $c=-{i \over 4}$, acting on the spinor
$\epsilon=\lambda 1 + \sigma e_1$.

\bea
\label{linsys}
\sigma \Omega_{1,2 \bar{1}} = \lambda \Omega_{\bar{1},12}= \lambda \Omega_{2,12} = \sigma \Omega_{2,2 \bar{1}}&=&0
\nn
\lambda \Omega_{1,12}+ \sqrt{2}i \sigma F_{12} &=&0
\nn
- \sqrt{2} i \lambda F_{\bar{1} 2} + \sigma \Omega_{\bar{1},2 \bar{1}} &=&0
\nn
i \lambda (-F_{1 \bar{1}}+F_{2 \bar{2}}) + \sqrt{2} \sigma \Omega_{\bar{2},2 \bar{1}} &=&0
\nn
\sqrt{2} \lambda \Omega_{\bar{2}, 12}- i \sigma  (F_{1 \bar{1}}+F_{2 \bar{2}}) &=&0
\nn
\partial_1 \lambda -{1 \over 2} \lambda (\Omega_{1,1 \bar{1}}+\Omega_{1,2 \bar{2}})
-{i \over \sqrt{2}} \sigma (F_{1 \bar{1}}+F_{2 \bar{2}})&=&0
\nn
\partial_1 \lambda + {1 \over 2} \lambda (\Omega_{1,1 \bar{1}}+\Omega_{1,2 \bar{2}})  &=&0
\nn
\partial_1 \sigma +{1 \over 2}\sigma(\Omega_{1, 1 \bar{1}}-\Omega_{1,2 \bar{2}}) &=&0
\nn
\partial_1 \sigma - {1 \over 2}\sigma(\Omega_{1, 1 \bar{1}}-\Omega_{1,2 \bar{2}}) +{i \over \sqrt{2}} \lambda
(F_{1 \bar{1}}-F_{2 \bar{2}}) &=&0
\nn
\partial_2 \lambda -{1 \over 2} \lambda (\Omega_{2,1 \bar{1}}+\Omega_{2, 2 \bar{2}}) &=&0
\nn
\partial_2 \lambda + {1 \over 2} \lambda (\Omega_{2,1 \bar{1}}+\Omega_{2, 2 \bar{2}}) 
-\sqrt{2} i \sigma F_{12} &=&0
\nn
\partial_2 \sigma +{1 \over 2} \sigma (\Omega_{2,1 \bar{1}}-\Omega_{2,2 \bar{2}}) &=&0
\nn
\partial_2 \sigma -{1 \over 2} \sigma (\Omega_{2,1 \bar{1}}-\Omega_{2,2 \bar{2}}) - \sqrt{2} i \lambda F_{\bar{1} 2} &=&0 \ .
\eea


\end{document}